\documentclass[aps,twocolumn,preprintnumbers,amsmath,amssymb]{revtex4}
\usepackage{epsfig}
\usepackage{amsmath}
\usepackage{graphicx}
\usepackage{dcolumn}
\usepackage{bm}
\usepackage{amssymb}
\usepackage{amsmath}
\usepackage{epsf}
\usepackage{subfigure}
\usepackage{epstopdf}
\usepackage{wrapfig}
\usepackage{pifont}
\usepackage{color}
\usepackage{wasysym}

\begin{document}

\newcommand{\pr}{\partial}
\newcommand{\rta}{\rightarrow}
\newcommand{\lta}{\leftarrow}
\newcommand{\ep}{\epsilon}
\newcommand{\ve}{\varepsilon}
\newcommand{\p}{\prime}
\newcommand{\om}{\omega}
\newcommand{\ra}{\rangle}
\newcommand{\la}{\langle}
\newcommand{\td}{\tilde}

\newcommand{\mo}{\mathcal{O}}
\newcommand{\ml}{\mathcal{L}}
\newcommand{\mathp}{\mathcal{P}}
\newcommand{\mq}{\mathcal{Q}}
\newcommand{\ms}{\mathcal{S}}

\newcommand{\nl}{$\newline$}
\newcommand{\nll}{$\newline\newline$}

\newcommand{\vspa}{\vspace{2mm}}
\newcommand{\vspb}{\vspace{3mm}}
\newcommand{\vspc}{\vspace{4mm}}

\title{Charge dynamics in an ideal cuprate $Ca_{2-x} Na_x Cu O_2 Cl_2$: optical conductivity from
Yang-Rice-Zhang ansatz}

\author{Navinder Singh}
\affiliation{Physical Research Laboratory, Navrangpura, Ahmedabad-380009 India.}

\begin{abstract}
We theoretically investigate charge dynamics in weakly coupled $CuO_2$ planes of the cuprate $Ca_{2-x} Na_x Cu
O_2 Cl_2$ (CNCOC) using Kubo formula for optical conductivity in the underdoped regime. The spectral function
needed in Kubo formula is obtained from  an analytical form of electron Green's function proposed (ansatz) by
Yang-Rice-Zhang (YRZ)  for the underdoped cuprates based on their previous renormalized mean field theory and on
the investigations of weakly coupled Hubbard ladders. Although to an unaided eye the results of the numerical
calculation look very similar to that found experimentally in [K. Waku etal. 2004\cite{waku}]  but a careful
examination with extended Drude formalism shows that YRZ ansatz for the calculation of optical
conductivity is not sufficient to understand the charge dynamics in $CuO_2$ planes of the cuprate CNCOC. More
physics is needed especially electromagnetic response from bound charges.
\end{abstract}

\maketitle

\section{Introduction}

Physical investigations at very small length scale (sub-atomic, atomic etc...) are usually indirect. To expose
or uncover the underlying physical reality one has to relay on indirect experimental probes. In simpler
situations few experimental probes can give one a consistent physical picture (for example spectral lines
of the atomic spectra suggested the discrete energy level structure of atoms). In more complex situations (as
for example in high temperature superconductors) one has to  deal with several experimental probes. Consistent
microscopic  physical picture of a material  can only be constructed if results of several experimental probes
are consistent with each other and comply with one {\it universal picture}. And that {\it universal picture} is
the major deriving force, and at the same time it also puts heavy demands on researchers of understanding an
arsenal of complex experimental probes (Fig.~\ref{ars}). In complex situations, understanding reached based
solely on few experimental results may not be consistent with other experimental results. One has to be careful.
 The situation is analogous to the famous story of blind men and an elephant. Each blind man ``feels'' different
part of the elephant and concludes that the elephant is like a snake (for a man who touches the trunk) or like a
pillar (for a man who feels the legs ) etc.  Only if they positively council (without fighting with each other)
they may reach on the ``universal'' picture of the elephant.
\begin{wrapfigure}{r}{0.2\textwidth}
  \vspace{-20pt}
  \begin{center}
    \includegraphics[width=0.2\textwidth]{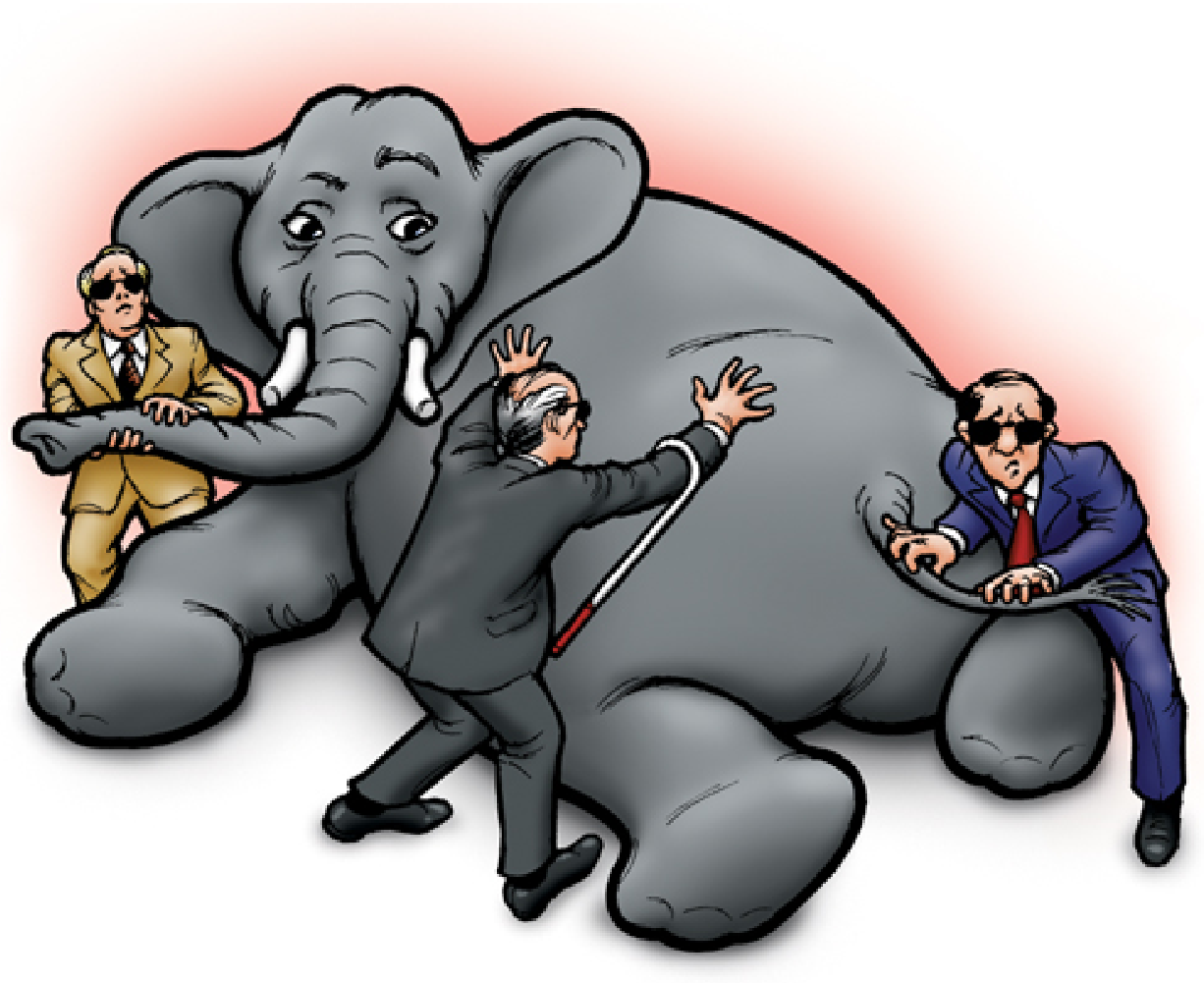}
  \end{center}
  \vspace{-15pt}
  \vspace{-10pt}
\end{wrapfigure}
\begin{figure}[h!]
\includegraphics[height = 5cm, width = 9cm]{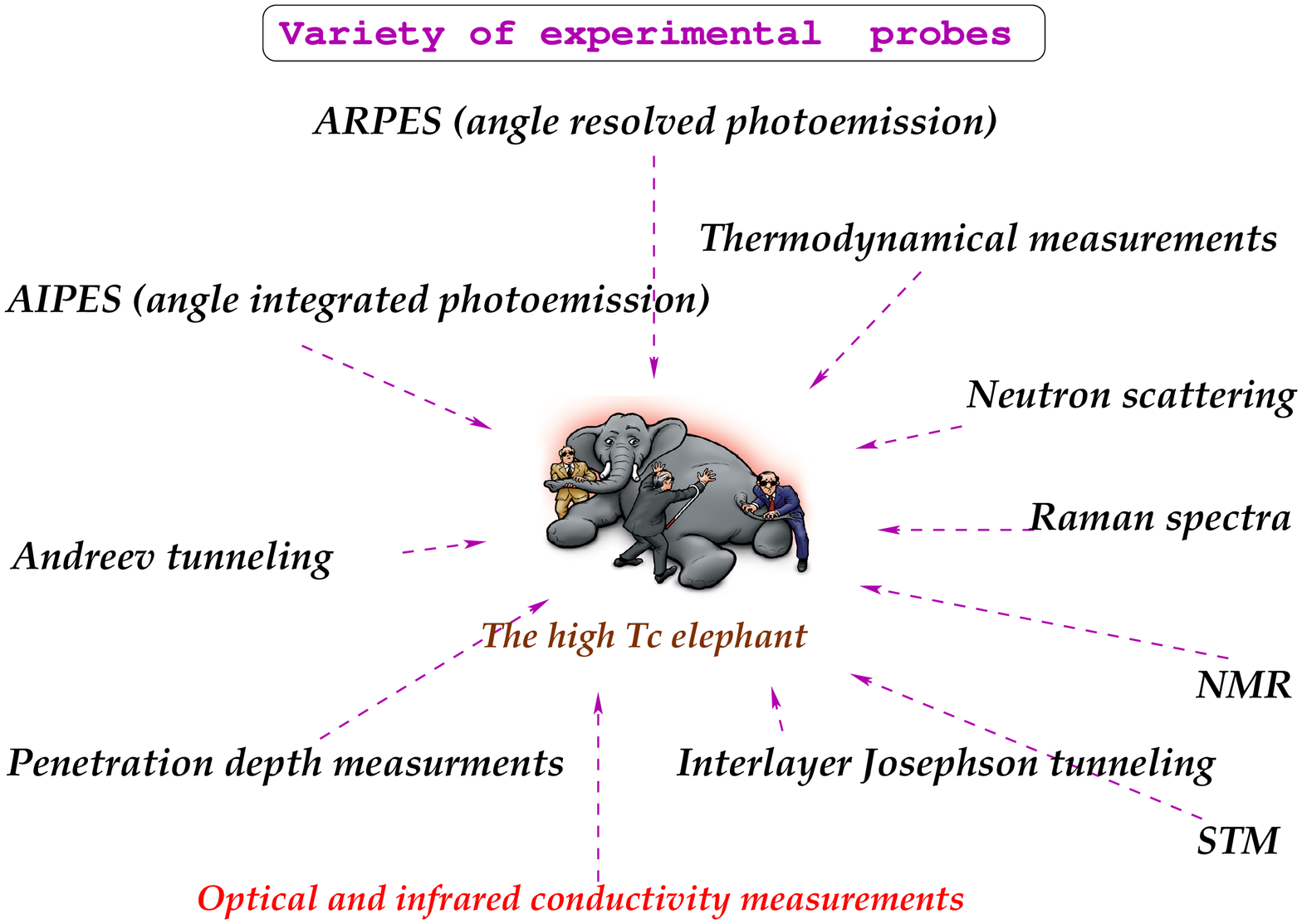}
\caption{An arsenal of complex experimental probes.}
\label{ars}
\end{figure}
At present, for the cuprate high temperature superconductivity problem we do not have a consistent {\it
universal picture} for the whole phase diagram consistent with all experimental probes (in any way comparable to
the description of ordinary superconductivity by Bardeen, Cooper, and Schrieffer). But there
are marvelous attempts\cite{ander1}. Cuprates are complex materials with many anomalous properties--not only
unconventional superconductivity but also unconventional normal state response. Some of the anomalous properties
are as follows. D.C. resistivity of hole doped cuprates just above the optimal doping is linear in temperature
$T$ (reminder: for a good metal obeying Fermi Liquid Theory (FLT), $\rho \sim T^2$, and $\rho\sim T^5$ if
phonons also contribute at $T$ ($T<< T_{Debye}$)). Away from optimal doping resistivity shows a very complex
behaviour\cite{dagotto}, and electron doped compounds do not show this linear in temperature behaviour of
resistivity! Hall coefficient is temperature dependent (for a good FLT metal it should be temperature
independent). For cuprates, Drude scattering rate turns out to be frequency and temperature dependent (for a
good FLT metal it is constant (again if phonons do not contribute in the temperature regime of interest)). In
the ``normal'' state above optimal doping real optical conductivity shows $\sim \frac{1}{\om}$ behaviour (in a
good FLT metal it is $\sim\frac{1}{\om^2}$ Drude behaviour). NMR relaxation rate shows substantial deviations
from linear in temperature behaviour ($T$-linear Korringa type). High $T_c$ cuprates has smaller coherence
length (roughly the size of the cooper pair) as compared to conventional (in accord with BCS theory)
superconductors. The cooper pair size turns out to be smaller than mean carrier-carrier separation, thus
mean-field type approximations in cuprate problem are questionable. There are many other anomalous properties
without proper understanding\cite{under}. Cuprates show variety of phases with temperature and doping and one of
the elusive phases is the pseudogap phase (Fig.~\ref{phases}). There are many views on pseudogap
phase\cite{yrzrev}, but recently, a broad picture at a phenomenological level of the pseudogap phase is proposed
for the cuprate problem. It is contained in Yang-Rice-Zhang (YRZ) ansatz\cite{yrzrev,yrzi}. This is based on
several inputs\cite{yrzrev}.

The key element of YRZ's phenomenological theory is an ansatz for the single electron propagator. This ansatz
is the outcome of series of investigations of the proposers over several decades and is primarily based
on their renormalized mean field theory\cite{rmft} and on the investigations of weakly coupled Hubbard
ladders\cite{ladders}. YRZ ansatz has been applied successfully to understand several anomalous properties of
the pseudogap phase\cite{yrzrev}. This has been applied with success to the interpretation of ARPES (angle
resolved photoemission) experiments on CNCOC\cite{arpes} and good agreement is seen between the calculated hole
Fermi pocket and experimental data\cite{yrzrev}. YRZ ansatz also correctly reproduce particle-hope asymmetry
seen in experiments of Yang etal\cite{yang}. This has also been applied to AIPES (angle integrated
photoemission) and qualitatively reproduce the key features of the spectra\cite{hash, yrzrev}. For STM (scanning
tunneling microscopy) spectra of constant quasi-particle energy contours on  BSCCO\cite{koh} one sees a
qualitative agreement with YRZ, although quantitative fits with experimental spectra are not
claimed\cite{yrzrev}. Raman Spectra is quite useful because it probes both the nodal ($B_{2g}$) and anti-nodal 
($B_{1g}$) charge dynamics\cite{ram}. Valenzuela and Bascones\cite{bas} found that YRZ qualitatively reproduce
features seen in the spectra\cite{tac} (in perticluar they deduce two energy scales, nodal and anti-nodal, with
opposite dependence on doping, nodal scale decreases with underdoping while the anti-nodal one increases). YRZ
has been applied to several other experimental results see for detail\cite{yrzrev}. In regard to the
microscopic picture, one should clearly distinguish between YRZ ansatz and preformed pair picture\cite{pre} in
which particle-hole symmetry is maintained.

Our interest here is in the optical spectra and in the behaviour of optical conductivity in the pseudogap
phase-- its variation with doping and temperature. Optical conductivity spectra of the cuprates is complex in
comparison to that of conventional superconductors. In conventional superconductors the gap is isotropic in
momentum and one gets a clear signature of the gap in the spectrum (In fact early pioneering experiments by
Tinkham etal\cite{tink} gave support to the energy gap model of superconductivity on which Bardeen highly relied
and later on a very successful theory of electromagnetic response was put forward by Mattis-Bardeen\cite{mb}).
In Cuprates gap(s) are anisotropic and optical conductivity spectra becomes intricate. Recently, experimentally
observed optical spectra of cuprates has been studied using the YRZ theory by Carbotte,  Nicol and
colleagues\cite{car}. Their investigations support the view that YRZ ansatz qualitatively reproduce low energy
behaviour of optical conductivity. 

Here, in the present investigation, we re-visit this problem. Our results do not bring any good news for
the applicability of YRZ ansatz to the low energy optical response of cuprates. We concentrated on a specific
compound $Ca_{2-x} Na_x Cu O_2 Cl_2$ (CNCOC). The results of our numerical calculation appears to be in
qualitative agreement with what has been found experimentally in\cite{waku} (to an unaided eye the results look
very similar to that found experimentally). But a careful examination with extended Drude formalism shows that
YRZ ansatz for the calculation of optical conductivity is not sufficient to understand the charge dynamics in
$CuO_2$ planes of CNCOC. {\it It seems that more physics is needed to fully understand the optical response.}

Rest of the paper is organized as follows. In the next subsection (subsection A) essential points of YRZ ansatz
and the cuprates are given. In subsection (B) a brief introduction to optical conductivity and Kubo formalism
is given. In section II, CNCOC system and experimental results are summarized. The calculation of optical
conductivity using YRZ ansatz and comparison with experiment is given in section III. In section IV extended
Drude model analysis is presented to point out the inconsistencies. We end with brief conclusion in section
V.

\begin{figure}
\includegraphics[height = 4.5cm, width = 8cm]{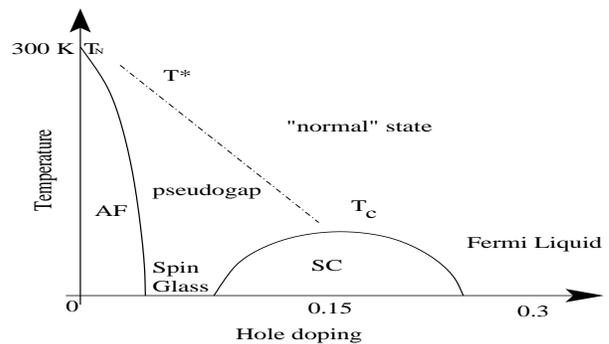}
\caption{Phase diagram of cuprates.}
\label{phases}
\end{figure}

\subsection{Essential points of YRZ ansatz and cuprates} 

The starting point of YRZ ansatz is the reasonably well understood regimes of extreme underdoping and overdoping
(Fig. 2). At overdoping one observes full Fermi surface and it disappears at zero doping. As one
shifts from overdoping to underdoping in the phase diagram Fermi surface evolves from full Fermi surface  to
only disconnected arcs (in specific directions in $k$-space) at underdoping to no Fermi surface at zero doping.
The key issue is: {\it how can one explain this doping dependent evolution of the Fermi surface using a
microscopic model?} The major problem is the intermediate regime and YRZ ansatz is an attempt to fill this gap.
As is well known, undoped cuprates are Mott insulators (more precisely charge transfer type) and it was pointed
out by Phil Anderson very early on that the operational elements in the cuprate superconductivity are the
$CuO_2$-planes. $Cu$ atoms in $CuO_2$ planes are in $d^9$ configuration (with one hole in the higher energy
anti-bounding $3d_{x^2-y^2}Cu--2p_{x(y)}$ orbital lying in $CuO_2$ plane). The holes localize (immobile) on
atomic sites due strong on-site coulomb repulsion. With further hole doping, new holes are created in $CuO_2$
planes. These new holes will not be in $Cu$ $d$-orbitals (because of strong Coulomb repulsion)  and tend to be
in oxygen $p$ orbitals. It was shown by Zhang and Rice\cite{zr} that if they form singlet pairs with the holes
in $Cu$ atoms then they will have lower energy. These singlet pairs are now called Zhang-Rice (ZR) singlets.
{\it We are interested in the transport properties of these ZR singlets.} This also reduces the cuprate problem
from three band to one band problem (although there are debates in the literature\cite{debates}). {\it The idea
is to reduce the problem to its bare essentials.} This motivates the famous $t-J$ model\cite{ander1} with no
double occupancy at the same site (the Hilbert space of the problem will not have any configuration in which any
site is doubly occupied--the projected Hilbert space). One can work in {\it unprojected} Hilbert space by using
the ideas of Renormalized Mean Field Theory (RMFT)\cite{ander1,yrzrev}. This is the first component of YRZ
ansatz. The second component is based on the analytical studies by Konik, Rice, Tsvelik, and Ludwig\cite{konik}
of 2-leg Hubbard ladders and a collection of weakly coupled Hubbard ladders. For weakly coupled 2-leg Hubbard
ladders the coherent part of Green's function turns out to be (treating the inter-ladder coupling in RPA):

\begin{eqnarray}
G(k_x,k_y,\om)&=& \frac{1}{G_0^{-1}(k_x,\om) - t_y(k_y)}\nonumber\\
G^0(k_x,\omega) &=& \frac{1}{\hbar\omega - \ep(k_x) -\frac{\Delta^2}{\hbar \omega +\ep(k_x)}}.
\label{konik}
\end{eqnarray}
Where $t_y$ is the transverse inter-ladder hopping. $\ep(k_y)$ and $\Delta$ are the bare band dispersion and
quasi-particle gap respectively. 
\begin{figure}[h!]
\includegraphics[height = 2cm, width = 6cm]{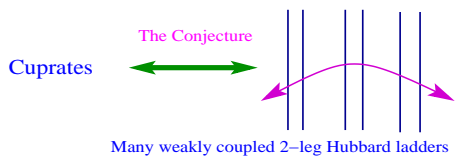}
\caption{YRZ conjucture: equivalence of many weakly coupled 2-leg Hubbard ladders and $CuO_2$ planes
in cuprates.}\label{ans}
\end{figure}

Based on the above two main elements, here is the key postulate: {\it continuous crossover from
the weak to strong interaction limit}. The coherent part of YRZ Green's function for the pseudogap state of
cuprates is postulated to be:
\begin{eqnarray}
G^{YRZ}(k,\omega) &=& \frac{g_t(x)}{\hbar \omega -\ep(k) -\Sigma_{pg}(k,\omega)},\nonumber\\
\Sigma_{pg}(k,\omega) &=&
\frac{|\Delta_{pg}(k)|^2}{\hbar\omega +\ep_0(k)}.
\label{yrz}
\end{eqnarray}
This is essentially based on intuition (Fig.~\ref{ans}). The dressed band dispersion $\ep(k)$ contains the
renormalization factors of RMFT.  Details of bare band and renormalized band dispersions are given in the next
section.

\subsection{Optical Conductivity}
Optical conductivity is the linear response function of an external A. C. electric field. In a rough classical
picture the charge carriers oscillate back-and-forth under the influence of an external A.C. electric field
(neglecting the weaker ($\frac{1}{c}$th) magnitude magnetic field effects). During this back-and-forth motion
they scatter by impurities and phonons causing the dissipation of energy (Joule heating). In the linear response
regime induced current density $\bf{J}(\bf{r},\omega)$ is related to the applied electric field
$\bf{E}(\bf{r'},\omega)$ by

\[\bf{J}(\bf{r},\omega) = \int \sigma(\bf{r},\bf{r'};\omega)\bf{E}(\bf{r'},\omega) d\bf{r'}.\]

If we are not in the anomalous skin effect regime (i.e., electric field does not  vary substantially on the
length scale of the mean-free-path of charge carriers) then $\sigma(\bf{r},\bf{r'};\omega) =
\sigma(\bf{r};\omega) \delta(\bf{r}-\bf{r'})$ and
\[\bf{J}(\bf{r},\omega) = \sigma(\bf{r};\omega)\bf{E}(\bf{r},\omega).\]
And further if the material is homogeneous:
\[\bf{J}(\bf{r},\omega) = \sigma(\omega)\bf{E}(\bf{r},\omega).\]
The optical conductivity (``Optical'', if $\omega$ is in optical frequency range) $\sigma(\omega)$ can be
calculated by using quantum mechanical expression for current density $J = \frac{\hbar}{2 m i}(\psi^\ast\nabla
\psi - \psi \nabla \psi^\ast)$ and it can be shown that
\begin{eqnarray}
&&\sigma(\omega,T)=\frac{2 \pi e^2}{\hbar V} \sum_{k_x,k_y,k_z} v_{o,x}^2(k_x,k_y)\times\\\label{kubo1}
&& \int_{-\infty}^{+\infty} dy \frac{f(y)-f(\omega+y)}{\omega}A(k_x,k_y,y) A(k_x,k_y,\omega+y).\nonumber\
\end{eqnarray}
This is called the Kubo formula (see for details\cite{mahan}). Here $V$ is the volume of the sample,
$v_{0,x}(k_x,k_y)=\frac{1}{\hbar} \frac{d \ep(k_x, k_y)}{d k_x}$ is the Fermi velocity, $f(\omega)
=\frac{1}{e^{\beta (\hbar\om -\mu_p)}+1} $ is Fermi-Dirac distribution function ($\beta = \frac{1}{k_B T}$,
$\mu_p$ is the chemical potential). Spectral function $A(k_x,k_y,\omega)$ is given by the usual expression  $
A(k_x,k_y,\omega) = - \frac{1}{\pi} Im G^{YRZ}(k_x,k_y, \omega+i 0^+)$. The coherent part of YRZ Green's
function is postulated, as given in equation (\ref{yrz}), thus the spectral function takes the form:
\begin{equation}
 A(k_x,k_y,\om) = \frac{\gamma}{ (\om -\frac{\ep(k_x,k_y)}{\hbar} - \frac{\Sigma_{pg}(k_x,k_y,\om)}{\hbar})^2 +
\gamma^2}
\label{spec}
\end{equation}
$\Sigma(k_x,k_y,\om)$ is given in equation (\ref{yrz}), and broadening $\gamma$ is introduced
phenomenologically\cite{car}. This gives finite life time of excitations due various scattering mechanisms
during transport. Renormalized dispersion $\ep(k_x,k_y,\om)$ and bare
band dispersion $\ep_0(k_x,k_y,\om)$ in the YRZ theory\cite{car,yrzrev} are
\begin{eqnarray}
\ep(k_x,k_y) &=& - 2 t(x) (\cos(k_x a) + \cos(k_y a)) \nonumber\\
&-& 4 t'(x) \cos(k_x a) \cos(k_y a) \nonumber\\
&-& 2 t''(x) (\cos(2 k_x a) +\cos(2 k_y a)) - \mu_p,
\end{eqnarray}
and $\ep_0(k_x,k_y,\om) = -2 t(x) (\cos(k_x a) + \cos(k_y a))$. Here $a$ is the lattice constant. In YRZ
theory\cite{yrzi,yrzrev} various band parameters are
\begin{eqnarray}
t(x)  &=& g_t(x)t_0 + \frac{3}{8} J \xi g_s(x),~~ \xi=0.338,~ J =\frac{t_0}{3},\nonumber\\
t'(x) &=& g_t(x) t'_0, ~~t'_0 = -0.3 t_0,~~g_t(x)=\frac{2 x}{1+x}\nonumber\\
t^{''}(x) &=& g_t(x) t_0^{''}  ~~ t_0^{''}  = 0.2 t_0,~~ g_s(x)=\frac{4}{(1+x)^2},
\label{para}
\end{eqnarray}
as calculated from the band structure of $Ca_2 CuO_2Cl_2$\cite{yrzi,matt}. $g_t(x)$ and $g_s(x)$ are
renormalization factors in YRZ theory. And the pseudogap $\Delta_{pg}(k_x,k_y) = \frac{1}{2} \Delta_{pg}^0(x)
(\cos(k_x a) -\cos(k_y a))$ with $\Delta_{pg}^0(x) = 0.6 t_0 (1-x/0.2)$ in $eV$. Usually $t_0 \sim 400 ~meV$ in
Cuprates.
Before we present the numerical calculation it is important to consider the experimental
results of the system under consideration.

\section{The  $Ca_{2-x} Na_x Cu O_2 Cl_2$ system and experimental results}
The $Ca_{2-x} Na_x Cu O_2 Cl_2$ (CNCOC) system is a near ideal cuprate with single $CuO_2$ plane per unit cell
and the coupling of $CuO_2$-planes in CNCOC is expected to be weaker than that of other cuprates like
$La_{2-x}Sr_x CuO_4$, $YBa_2Cu_3O_7$, $Bi_2Sr_2CaCu_2O_8$ etc. This is due to the fact that in CNCOC the apical
atoms are Chlorine ions (instead of Oxygen as in other cuprates) and Chlorine planes have more ionic character.
This exert pulling of the d-electron on $Cu$ ion towards the apical axis and thereby leads to very pointed
octahedron (more positive charge on the $Cu$ ion pull the planner Oxygen atoms towards itself and thus pushing
apical atoms further out and creates a pointed octahedron). Thus the c-axis distance in CNCOC is about $15~\AA$,
larger as compared to that in other cuprates ($\sim 10~ \AA$). Also CNCOC has no orthorhombic distortion from
tetragonal structure as it is cooled through the pseudogap boundary. Due to these qualities CNCOC can be
regarded as a {\it better} cuprate regarding charge dynamics in $CuO_2$ planes.

\begin{figure}[h!]
\includegraphics[height = 3cm, width = 8cm]{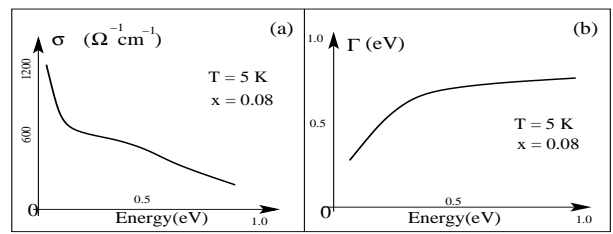}
\caption{Schematic view of the experimental results (see for original\cite{waku})}\label{sche}
\end{figure}

Charge dynamics of CNCOC has been measured experimentally in a beautiful piece of work by Waku etal.\cite{waku}.
They measure D.C. resistivity and A.C. optical conductivity of single crystals of CNCOC grown by flux
method\cite{waku}. They also measure temperature dependence of low energy $(\hbar \om \lesssim 1~ eV)$ optical
conductivity at various doping levels (Mott gap appears at higher energy $\hbar\om \simeq 2~ eV$ thus at $(\hbar
\om \lesssim 1 eV)$ one is in low energy intraband response regime). The conductivity obtained is shown in
Fig.(8) of their paper\cite{waku} (it is also shown schematically in Fig.~\ref{sche}(a) in the present
manuscript). It is well known that spectrum below $1 ~eV$ in the cuprates cannot be fitted with single Drude
model. This is also the case with CNCOC. It is this part of the spectrum that we will consider in our study.  In
their experimental study\cite{waku}, they analyzed the experimental results with both two-component Drude model
and generalized or extended Drude model (also called memory function formalism\cite{basov}). We consider here
their extended Drude model analysis of the experimental data.  In this model all the spectrum below $1~eV$ is
assigned to an itinerant state, with scattering rate and effective mass of charge carriers having frequency
dependence. In this model the complex optical conductivity can be written as:
\begin{equation}
\tilde{\sigma}(\om) = \frac{\om^2_p}{4\pi}\frac{1}{1/\tau(\om) - i \om m^\ast(\om)},~~\om_p^2 = \frac{4 \pi n
e^2}{m^\ast}.
\label{gdm}
\end{equation}
Here the scattering rate $(\Gamma(\om) = \frac{1}{\tau(\om)})$ and effective mass $m^\ast(\om)$ both have
frequency dependence. By fitting this with their experimental results they deduced frequency dependence of
$\Gamma(\om)$. They found that at low frequencies $\hbar \om \lesssim 0.4 ~eV$ scattering rate is almost
proportional to $\om$ ($\Gamma(\om) = \Gamma_0 + C \om$). $C$ turns out to be almost temperature independent
and  $\Gamma_0$ increases with increasing temperature (ref. figure (10) in their paper\cite{waku}). They found
that above $\hbar \om \gtrsim 0.4~eV$ the scattering rate saturates to a constant value (very weakly dependent
on temperature and doping). This is also shown schematically in Fig.~\ref{sche}(b).  They remark that this
saturation behavior is similar to resistivity saturation.
{\it In the next section we will see that although the optical conductivity as calculated numerically using YRZ
theory below $1~eV$ appears qualitatively (to an unaided eye) in agreement with the experimental results
(including the magnitude of conductivity (few hundred $\Omega^{-1} cm^{-1}$)) {\bf but the bump in
conductivity due to the presence of pseudogap shifts to lower energy regime with increasing  doping
$\Delta_{pg}^0(x) = 0.6 t_0 (1-x/0.2)$ in the numerical calculation using YRZ theory and this is
also reflected in the scattering rates. No doping independent saturation in $\Gamma(\om)$ is observed in the
numerical study. But in the experiments this kind of doping independent saturation cut-off at $\hbar \om
\simeq 0.4 ~eV$ in $\Gamma(\om)$ is observed}. These are the main results of the present investigation and this
is studied in greater detail in the next two sections.}

\section{Optical conductivity from YRZ ansatz and comparison with experiment}

Our aim in this section is to numerically compute optical conductivity  using YRZ ansatz for CNCOC
system and to check how does it compare with the experiment.
\begin{figure}[h!]
\centering
\begin{tabular}{cc}
\includegraphics[height = 3cm, width =4.3cm]{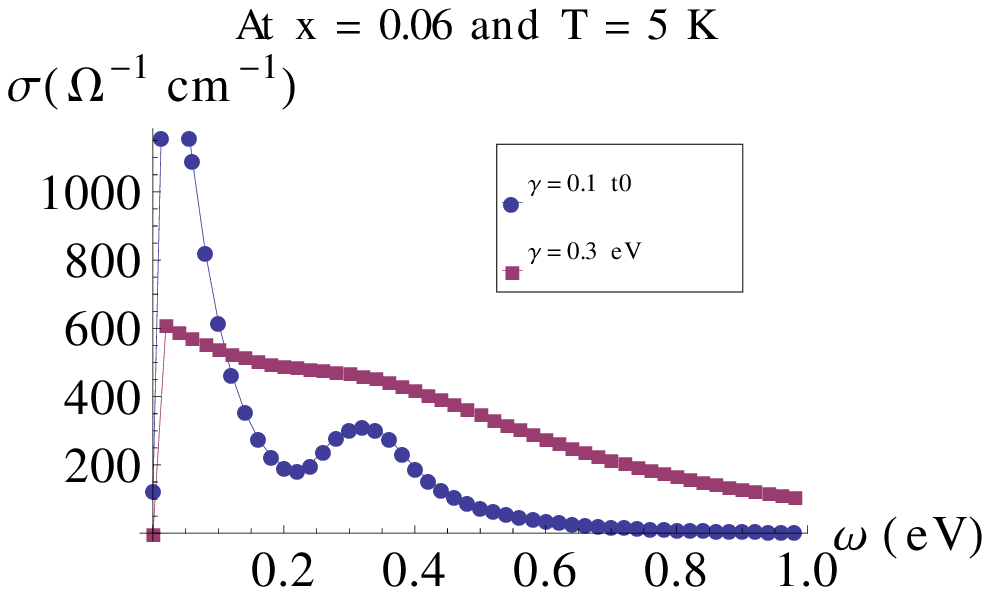}&
\includegraphics[height = 3cm, width =4.3cm]{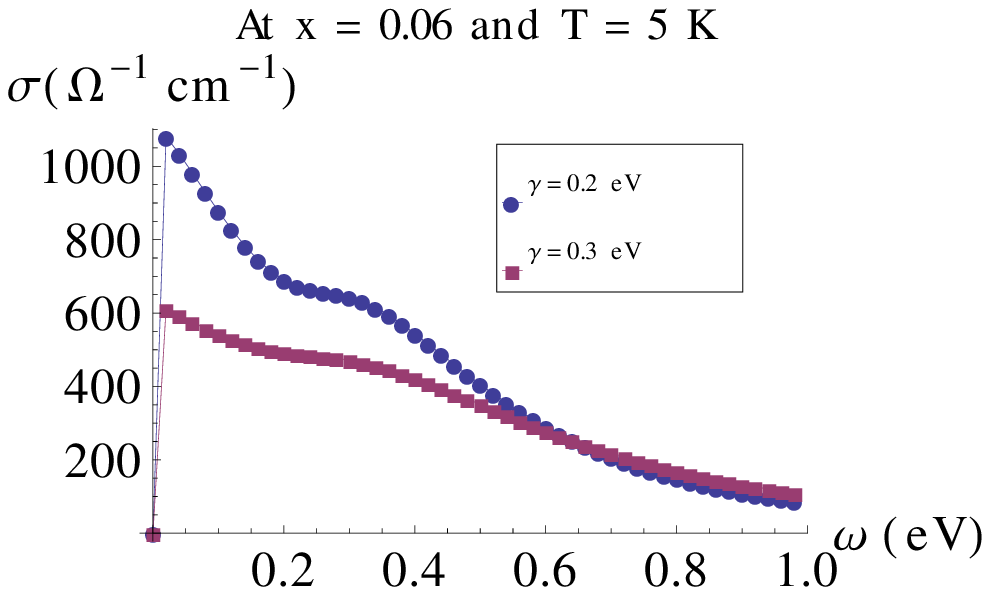}\\
(a)&
(b)\\
\includegraphics[height = 3cm, width =4.3cm]{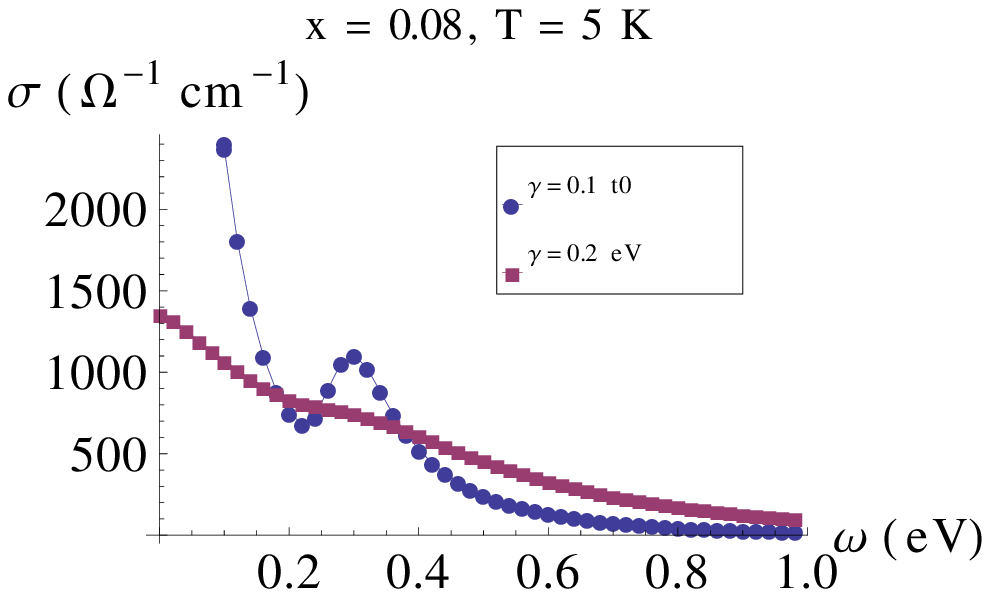}&
\includegraphics[height = 3cm, width =4.3cm]{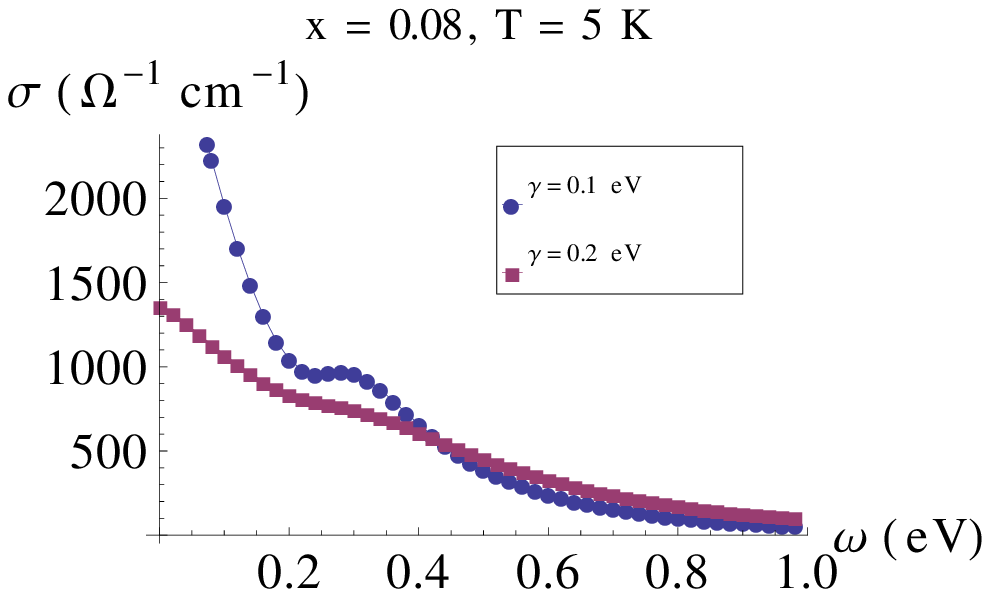}\\
(c)&
(d)\\
\includegraphics[height = 3cm, width =4.3cm]{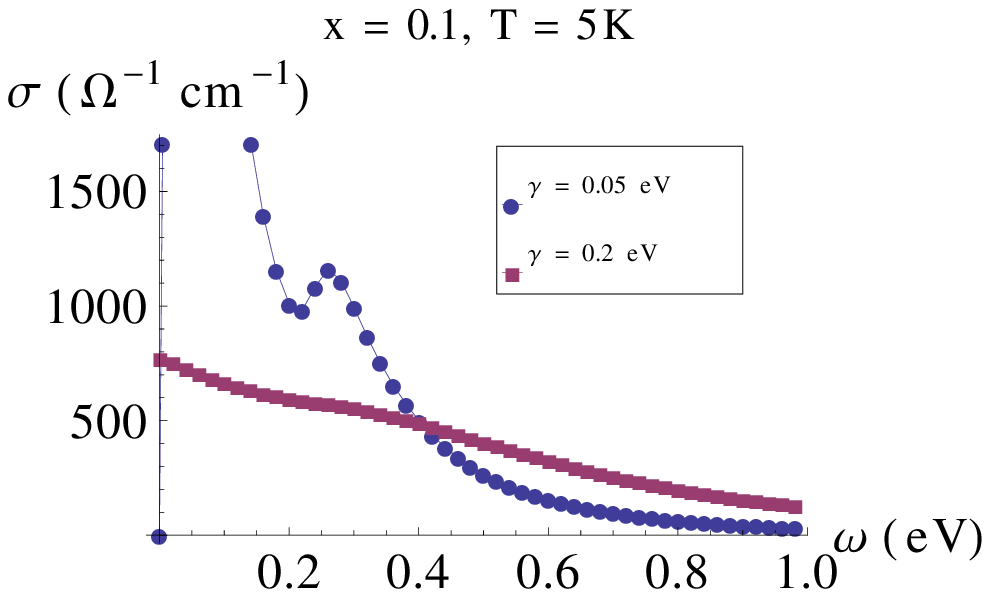}&
\includegraphics[height = 3cm, width =4.3cm]{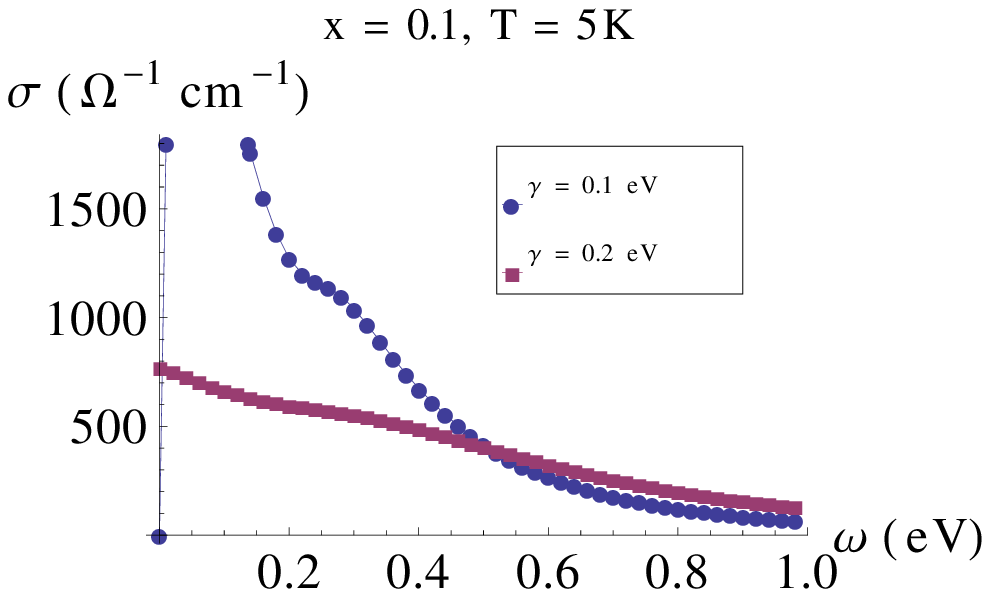}\\
(e)&
(f)\\
\end{tabular}
\caption{(a) Optical conductivity (in $\Omega^{-1} cm^{-1}$) vs frequency (in $eV$) at
$x = 0.06$ and at temperature $T = 5 K$. Comparison at different $\gamma$ but with psuedogap. The value of the
Drude scattering rate is given in the legend. All the energy scales are measured in $eV$. Figures in the second
and 3rd row are for $x=0.08$ and $x=0.1$ respectively.  In all figures $t_0 = 0.45 eV$.}
\label{set1}
\end{figure}
For the computation we needed a number of parameters of  CNCOC system.  These are taken from the
literature\cite{yrzi,matt,car} and tabulated below (Table I).

\begin{table}[h!]
\caption{Various parameters for CNCOC system.}
\begin{tabular}{|p{2cm}|p{2cm}|p{2cm}|p{2cm}|}
\hline
$CuO$-bond \ a-axis& b-axis & c-axis & $t_0$\\
\hline
$\simeq 3.87~\AA$  & $\simeq 3.87~\AA$  &  $\simeq 15~\AA$ &  $\simeq 0.45~eV$\\
\hline\hline

$\mu_p$ at $x=0.06$& $\mu_p$ at $x=0.08$ & $\mu_p$ at $x=0.1$\\
\hline

$\simeq -0.2~eV$ & $\simeq -0.23~eV$ & $\simeq-0.25~eV$\\
\hline
\end{tabular}
\end{table}

The superexchange constant is $J\simeq 0.13 ~eV$ for $Sr_2CuO_2Cl_2$ and for most of the cuprates $t_0$ is in
$0.3~eV \lesssim t_0 \lesssim 0.5~ eV$. Here we take $t_0 = 0.45~eV$\cite{car}. At $x=0.05$, $\Delta_{pg}^0 =
180~meV$, ARPES do detect this pseudogap ($\Delta^{ARPES}_{pg} \simeq 200~meV$) at this doping around the point
$(\pi,0)$ in the brillion zone and also co-existing nodal metal\cite{arpes}.

For numerical computation we considered a 3-D sample of $100\times100\times100$ lattice points (thus with
sample length, width, and hight: $100 a$, $100 a$, and $100 c$ (where $a$ is the $a$-axis ($a=b$) and $c$ is
the $c$-axis lattice constant)). This size of the sample is sufficiently large (as has been numerically
verified) and size effects can be neglected. After performing the sum over $k_z$ in equation (3) we
obtain
\begin{eqnarray}
&&\sigma(\omega,T)=\frac{4 \pi e^2 \hbar}{N^2 c a^2} \sum_{k_x=-\frac{\pi}{a},k_y =
-\frac{\pi}{a}}^{+\frac{\pi}{a},+\frac{\pi}{a}} v_{o,x}^2(k_x,k_y)\times\\
&& \int_{-\infty}^{+\infty} dy \frac{f(y)-f(\omega+y)}{\omega}A(k_x,k_y,y) A(k_x,k_y,\omega+y).\nonumber\
\label{kubo2}
\end{eqnarray}
Here we have redefined the units, now the frequency $y,~\om$ is measured in $eV$ and $\{k_x,k_y\} =
-\frac{\pi}{a}, -\frac{\pi}{a} + \delta,...,+\frac{\pi}{a}$, with $\delta = \frac{\pi}{N a}$, $N = 100$. And the
spectral function is
\begin{equation}
A(k_x,k_y,\om) = \frac{\gamma}{(\om -\ep(k_x,k_y) - \Sigma_{pg}(k_x,k_y,\om))^2 +\gamma^2}.
\label{spec2}
\end{equation}
Here $\gamma$ and $\om$ measured in energy units (in $eV$). Band dispersion and self energy is also measured in
$eV$ and conductivity has the right units $\Omega^{-1} cm^{-1}$.

Numerical computation is done on Mathematica-$8$ numerically by writing a small program.

\begin{figure}[h!]
\centering
\begin{tabular}{cc}
\includegraphics[height = 3cm, width =4.2cm]{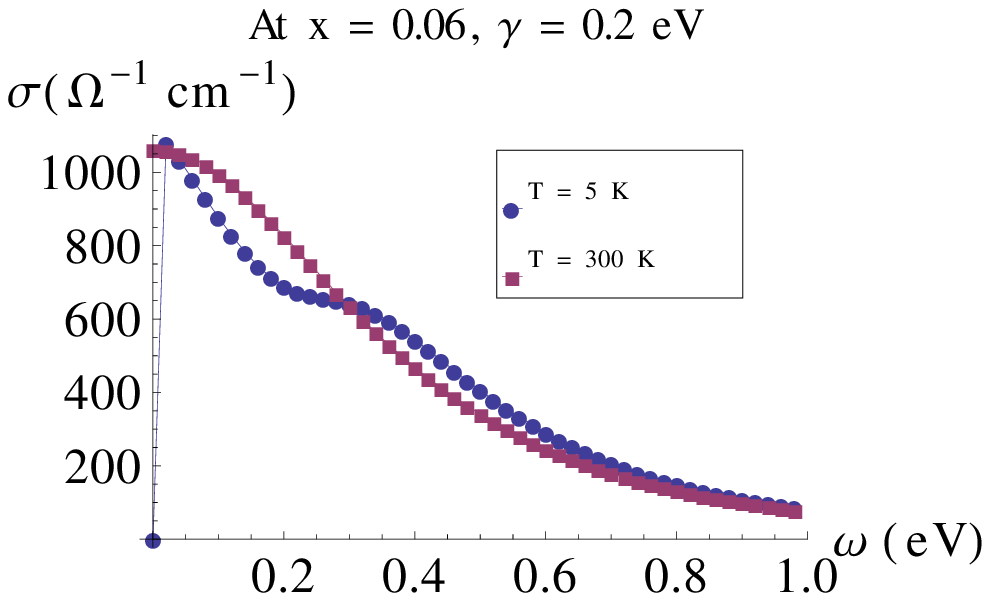}&
\includegraphics[height = 3cm, width =4.2cm]{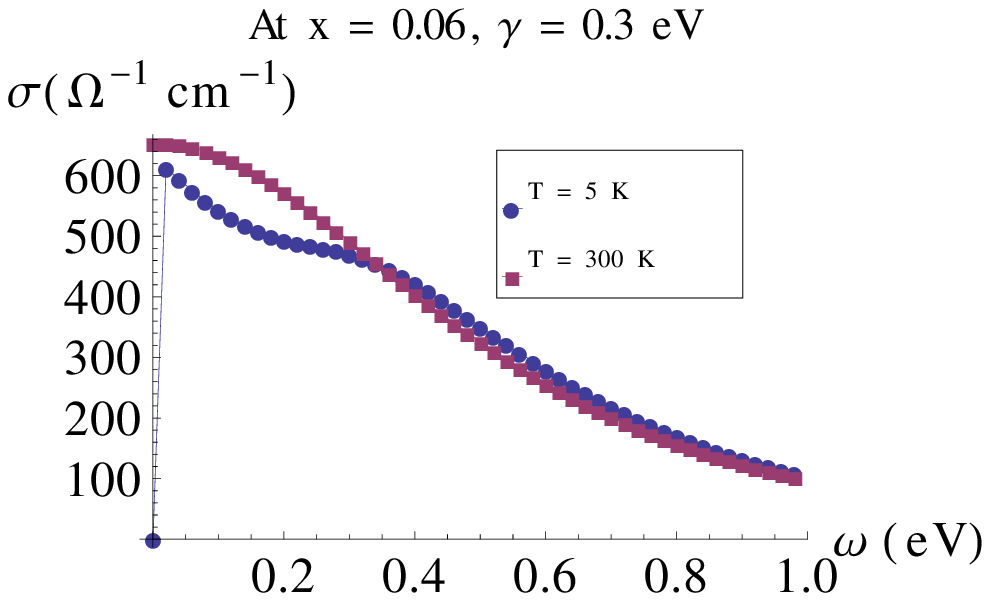}\\
(a)&
(b)\\
\includegraphics[height = 3cm, width =4.2cm]{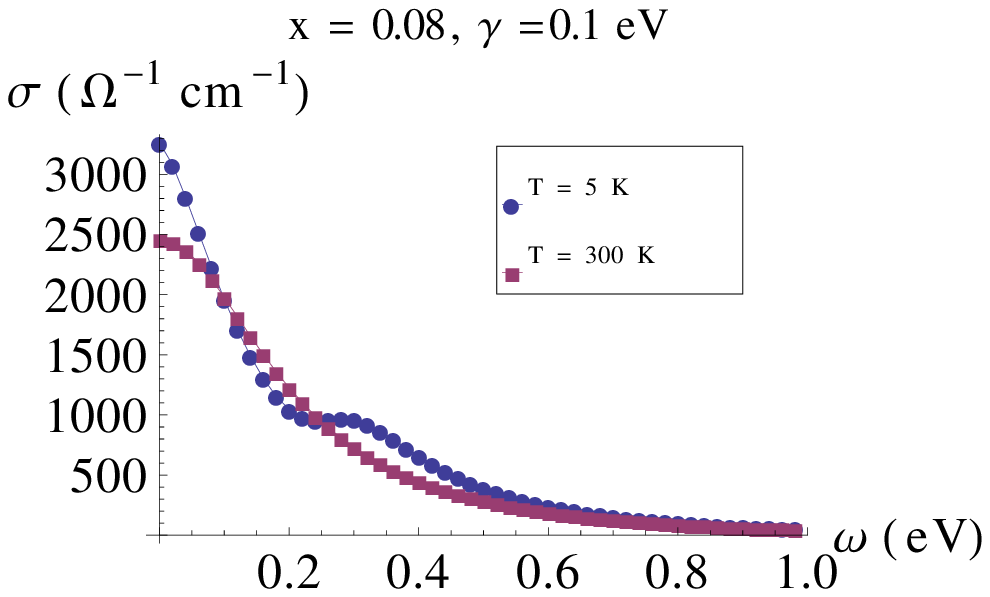}&
\includegraphics[height = 3cm, width =4.2cm]{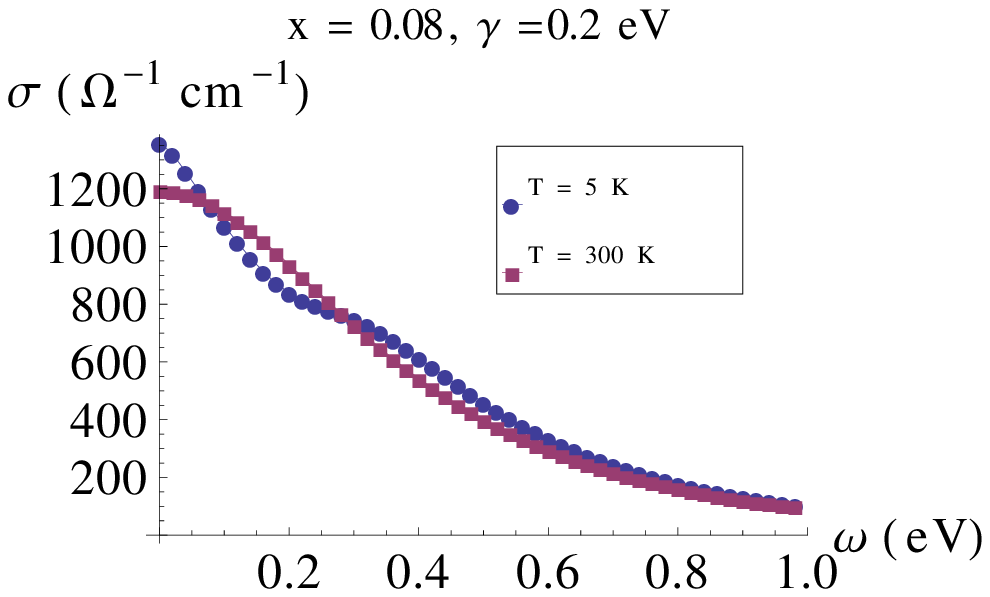}\\
(c)&
(d)\\
\end{tabular}
\caption{(a) With and without pseudogap at $x=0.06$ and $\gamma = 0.2~eV$. Other plots are at different
values of $x$ and $\gamma$ as written in plot titles.}
\label{set2}
\end{figure}

In Fig.~(\ref{set1}(a)) we plot optical conductivity as a function of frequency at two different $\gamma$'s and
at temperature $T = 5~K$ and hole doping $x=0.06$. Dip seen at around $0.2~eV$ is a signature of pseudogap which
broadens with increasing $\gamma$ (red solid squares for $\gamma= 0.3 ~eV$ and blue filled circles for $\gamma =
0.1 t_0~eV$). This pseudogap signature (Dip at around $0.2~eV$) disappears at higher temperature. This is shown
in Fig.~(\ref{set2}(a)) where $T=300~K$ and we do not have pseudogap at this temperature. {\it The conductivity
obtained at $\gamma =0.3 ~eV$  resembles closely what has been experimentally found (Fig.(8) of Waku
etal.\cite{waku})}. We will see in the next section that although it appears qualitatively in agreement with
what has been found experimentally but careful examination shows inconsistencies with YRZ. The other figures are
plotted for $x=0.08$ and $x=0.1$ as written in the figure titles.
\begin{figure}[!ht]
\centering
\begin{tabular}{cc}
\includegraphics[height = 3cm, width =4.2cm]{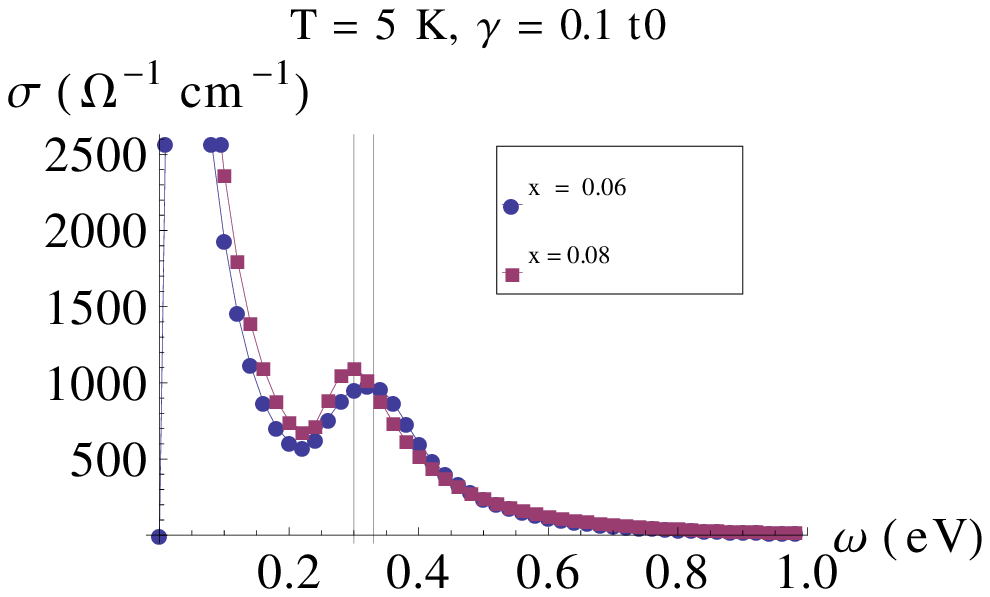}&
\includegraphics[height = 3cm, width =4.2cm]{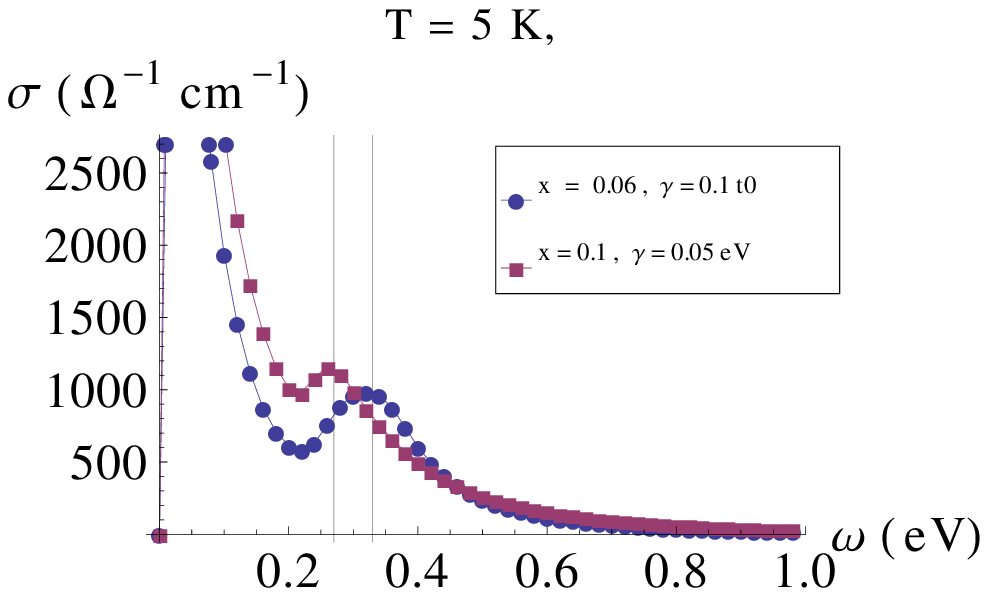}\\
(a)&
(b)\\
\end{tabular}
\caption{With increasing doping $x$, size of the pseudogap decreses $\Delta^0_{pg}(x) = 0.6 t_0 (1-x/0.2)$.
This should result in leftward shifting of the gap signature in the optical conductivity. Figure (a) show this
shifting for $x_1=0.06$ and $x_2 = 0.08$. Shift in $\Delta^0_{pg}(x)$ is $\delta\Delta^0_{pg}(x) =  3 t_0
(x_2-x_1) \simeq 0.03 eV$. This roughly agrees with shift in the graph (the distance between x-axis ticks is
$0.05 eV$). (b) Shifting for $x=0.06$ and $x =0.1$. Here also $\delta\Delta^0_{pg}(x) =  3 t_0 (x_2-x_1) \simeq
0.054$. This also roughly agrees with shift in the graph.}
\label{set3}
\end{figure}

In Fig.~(\ref{set3}(a)) we plot optical conductivity as a function of frequency at two different doping levels
($x=0.06$, and $x=0.08$) and at temperature $T = 5~K$ and at $\gamma = 0.1 t_0$. There is a clear shifting
of the pseudogap bump to lower frequencies with increasing doping $x$.  This is consistent with
decreasing $\Delta^0_{pg}(x) = 0.6 t_0 (1-x/0.2)$ with increasing $x$. In Fig.~(\ref{set3}(b)) the same
shifting is shown for $x=0.06$ and $x=0.1$. Shifting seen in the graph agrees with $\delta\Delta^0_{pg}(x) =  3
t_0 (x_2-x_1) \simeq 0.054$ as explained in figure caption. This shifting of pseudogap signature in
conductivity results similar shifting of the signature of pseudogap in the generalized Drude model scattering
rate. But, in contrast, this kind of shifting is not seen in the experiment of Waku etal\cite{waku}. This is the
subject of the next section.

\section{Extended Drude model analysis and inconsistencies with YRZ model}

In this section we will show that optical conductivity as computed using YRZ is inconsistent with what has been
experimentally observed in\cite{waku}. {\it We will deduce the above statement by the method of {\it reductio ad
absurdum} of logic, i.e., proof by contradiction. So let us assume that YRZ is the correct model for the
computation of optical conductivity in the low frequency regime which we are considering, i.e., the
underlying transport properties of quasi-particles are captured by YRZ. Now, as done in the experimental paper
by Waku etal.\cite{waku} we analyze the optical conductivity from YRZ ansatz with the extended Drude model and
extract the frequency dependent scattering rate. If the scattering rate so deduced agrees with scattering rate
deduced with similar analysis of the experimental data, then, YRZ ansatz is consistent with what has been
experimentally observed, otherwise,it is not.}

Now, as mentioned before (below equation (\ref{gdm})), they\cite{waku} deduce frequency dependence of the
scattering rate  $\Gamma(\om)$ of extended Drude model by using their experimental data. They found that at
low frequencies $\hbar \om \lesssim 0.4 ~eV$ scattering rate is almost proportional to $\om$ ($\Gamma(\om) =
\Gamma_0 + C \om$) and $C$ turns out to be almost temperature independent. And $\Gamma_0$ increases with
increasing temperature (refer to figure (10) in their paper\cite{waku}). They see that above $\hbar \om \gtrsim
0.4~eV$ the scattering rate saturates to a constant value. 

\begin{figure}[!ht]
\centering
\begin{tabular}{cc}
\includegraphics[height = 3cm, width =4.2cm]{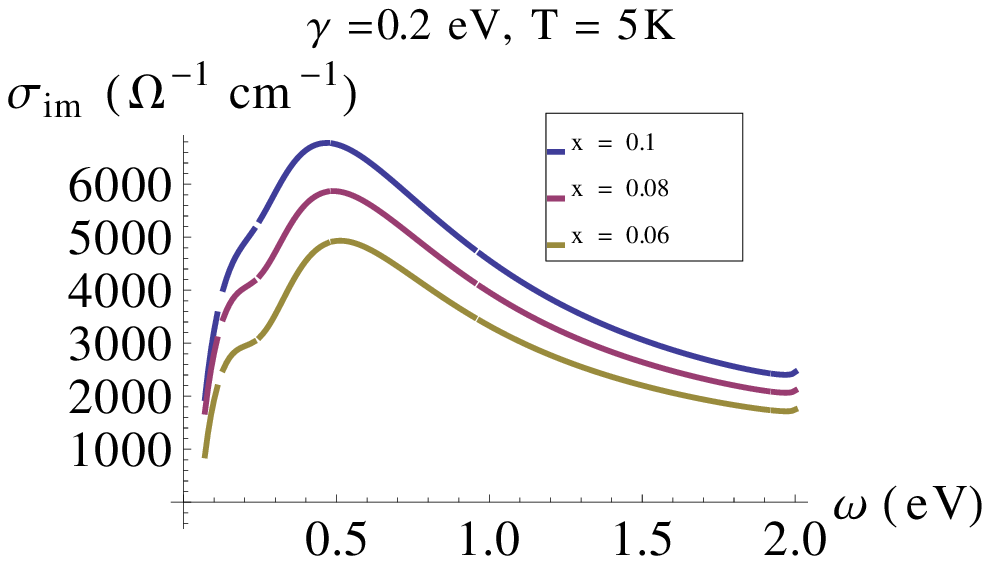}&
\includegraphics[height = 3cm, width =4.2cm]{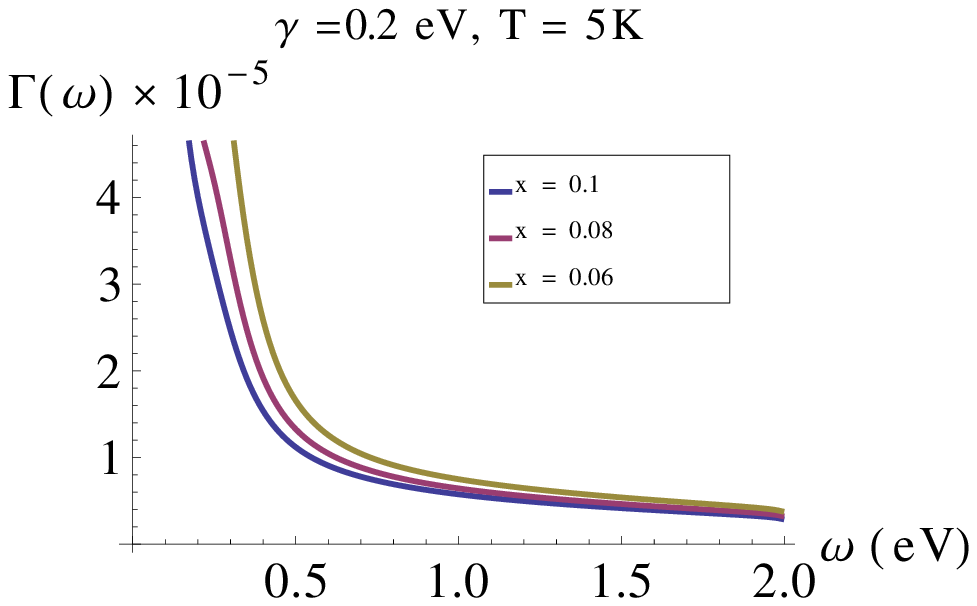}\\
(a)&
(b)\\
\end{tabular}
\caption{(a) Imaginary part of conductivity $\sigma_{im}(\omega)$ calculated from the real part by
Kramers-Kronig transformation. The scattering rate $\gamma$ and temperature $T$ is fixed at $0.2 eV$ and $5 K$
in these two graphs. The calculation is done for hole dopings $x=0.06$, $x=0.08$, and $x=0.1$. We see that
higher the doping higher is the conductivity and the peak in conductivity shifts to lower frequencies with
increasing doping as is investigated in figure~(\ref{set3}). The scattering rate $\Gamma(\om)$ obtained from
extended Drude model monotonically decrease with increasing frequency. This is in sharp contrast with the
experimental observations of Waku et al\cite{waku} where $\Gamma(\om)$ first increases linearly with frequency
and then saturates at around $\omega_c \sim 0.4 eV$.}
\label{set4}
\end{figure}

In our case, the scattering rate can be computed from\cite{basov}:
\begin{eqnarray}
\Gamma_{gd}(\om)&\equiv&\frac{1}{\tau(\om)} = \frac{n e^2}{m^\ast}
Re\left(\frac{1}{\tilde{\sigma}(\om)}\right)\nonumber\\
& = &\frac{n e^2}{m^\ast}\frac{\sigma(\om)}{(\sigma(\om))^2
+ (\sigma_{im}(\om))^2}.
\end{eqnarray}
For this we need to compute imaginary part of conductivity $\sigma_{im}(\om)$ (real part is $\sigma(\om)$). As
we are dealing with the casual perturbation and response relationship, the real and imaginary parts are related
by Kramers-Kronig relations.  By doing a numerical Kramers-Kronig inversion we obtain $\sigma_{im}(\om)$ (as
plotted in Fig.~(\ref{set4}(a))) and then we calculate $\Gamma_{gd}$. In Fig.~(\ref{set4}(b)) we plot
$\Gamma(\om) =\frac{\sigma(\om)}{(\sigma(\om))^2
+ (\sigma_{im}(\om))^2} $; ($\Gamma_{gd}(\om) = \frac{n e^2}{m^\ast} \Gamma(\om)$) at $\gamma =0.2 ~eV$.

\begin{figure}[!ht]
\centering
\begin{tabular}{cc}
\includegraphics[height = 3cm, width =4.2cm]{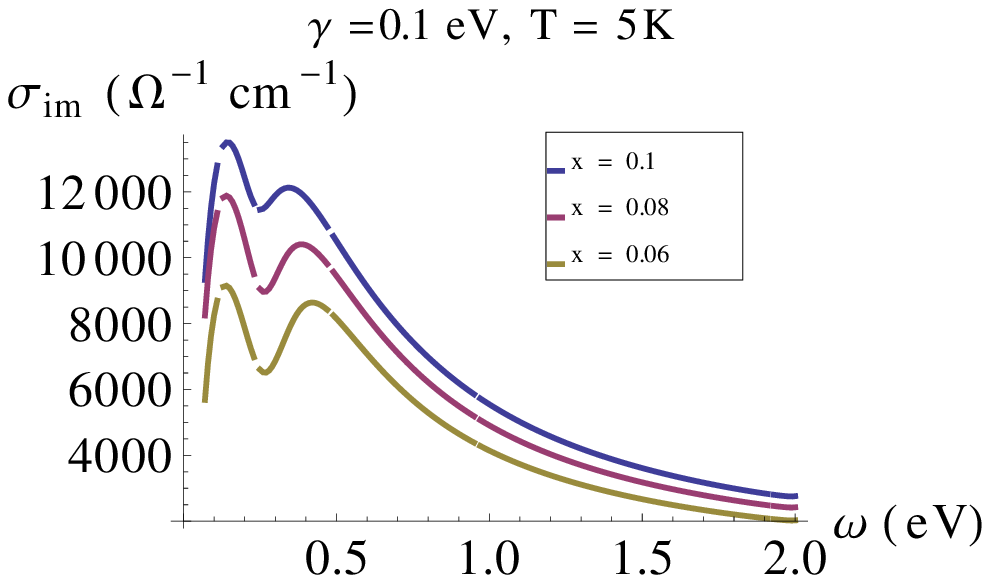}&
\includegraphics[height = 3cm, width =4.2cm]{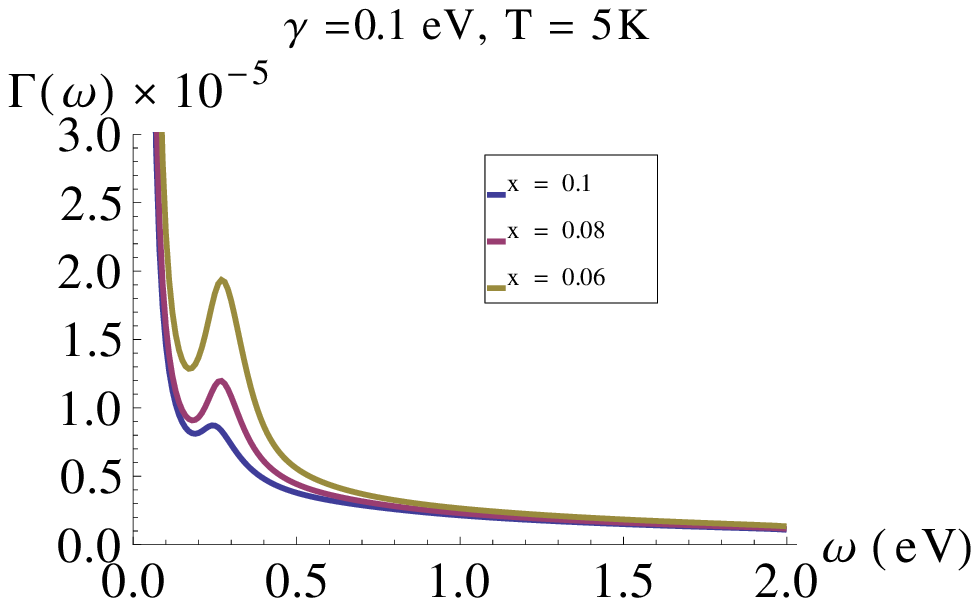}\\
(a)&
(b)\\
\end{tabular}
\caption{The similar analysis as in figure 8 with $\gamma =0.1$.}
\label{set5}
\end{figure}

We see clearly from Fig.~(\ref{set4}(b)) that the scattering rate $\Gamma(\om)$ obtained from
extended Drude model monotonically decrease with increasing frequency. This is in sharp contrast with the
experimental observations of Waku et al\cite{waku} where $\Gamma(\om)$ first increases linearly with frequency
and
then saturates at around $\omega_c \sim 0.4 eV$. Also with reduced doping $\Gamma(\om)$ shifts to lower
$\hbar\om$ in numerical study. Comparison of Fig.~\ref{sche}(b) (or figure (10) in\cite{waku}) with
Fig.~\ref{set4}(b) clearly points out the inconsistencies.  The smeared out features of the pseudogap at higher
$\gamma$ ($\gamma =0.2 ~eV$ in Fig.~(\ref{set4})) can be seen at lower $\gamma$ ($\gamma =0.1~eV$ in
Fig.~(\ref{set5})). But the pseudogap peak in $\Gamma(\om)$ shifts to higher $\hbar\om$ with reduced doping
(similar to the shift of $\sim 0.05~eV$ between the peaks at $x=0.06$ and $x=0.1$ in Fig.~\ref{set3}(b)). Also
the behaviour of two scattering rates ( Fig.~\ref{sche}(b) and Fig.~\ref{set5}(b)) is qualitatively different.

\section{Conclusion}
We theoretically investigated charge dynamics in weakly coupled $CuO_2$ planes of the cuprate $Ca_{2-x} Na_x Cu
O_2 Cl_2$ (CNCOC) using YRZ ansatz for the single particle Green's function in the pseudogap state. To an
unaided eye, the results of our numerical calculation appears to be in qualitative agreement with what has been
found experimentally in\cite{waku}. But a careful examination with extended Drude formalism shows that YRZ
ansatz for the calculation of optical conductivity is not sufficient to understand the optical conductivity in
$CuO_2$ planes of the compound CNCOC. It seems that more physics is needed to fully understand the optical
response especially the response from bound charges.

\end{document}